\newif\ifletter
\lettertrue
\letterfalse

\ifletter
\documentclass[showpacs,showkeys,amsmath,amssymb,superscript,twocolumn]{revtex4-1}
\else
\documentclass[a4paper,11pt]{article}
\usepackage{authblk}
\fi

\usepackage{fourier}
\usepackage{amsopn}
\usepackage{amsmath}
\usepackage{amssymb}
\usepackage{hyperref}
\usepackage{graphicx}
\usepackage{color}
\usepackage{listings}
\usepackage{todonotes}
\usepackage{tikz}
\usepackage{subfig}
\usetikzlibrary{arrows,decorations.pathreplacing}
\usetikzlibrary{arrows,automata,matrix}
\usepackage[ruled,vlined]{algorithm2e}
\usepackage{pgffor}
\usepackage{changepage}      

\newcommand{\eg}{{\emph{e.g.\/}}}
\newcommand{\ie}{{\emph{i.e.\/}}}
\newcommand{\Complex}{\ensuremath{\mathbb{C}}}

\newcommand{\XX}{\ensuremath{\mathcal{X}}}

\newcommand{\ket}[1]{\ensuremath{|#1\rangle}}
\newcommand{\bra}[1]{\ensuremath{\langle#1|}}
\newcommand{\ketbra}[2]{\ensuremath{\ket{#1}\bra{#2}}}
\newcommand{\braket}[2]{\ensuremath{\langle{#1}|{#2}\rangle}}

\newcommand{\ceil}[1]{\ensuremath{\left\lceil #1 \right\rceil}}
\newcommand{\1}{{\rm 1\hspace{-0.9mm}l}}

\newtheorem{theorem}{Theorem}
\newtheorem{corollary}{Corollary}

\newtheorem{fact}{Fact}

\def\papertitle{Quantum search with prior knowledge}
\ifletter\else
\title{\papertitle}

\author{Przemys{\l}aw Sadowski\thanks{psadowski@iitis.pl}\;}
\affil{Institute of Theoretical and Applied Informatics, Polish Academy of 
Sciences, Ba{\l}tycka 5, 44-100 Gliwice, Poland}
\date{}
\fi

\begin{document}


\ifletter\else
\maketitle
\fi

\begin{abstract}
The aim of this work is to develop a framework for realising quantum network
algorithms with the use of prior knowledge about the structure of the network.
We seek to obtain computational methods that allows us to locally determine
network properties in a quantum superposition and drive the walk behaviour
accordingly. In particular, we consider a network that consists of different
types of edges, such that the transitions between nodes result in extra
edge-dependent phase shift. We combine amplitude amplification and phase
estimation to develop an algorithm for exploring such networks. In the layered
neural network inspired case we obtain linear increase of the search complexity
with exponential growth of the nodes number. We show that in consequence one is
able to perform quantum search algorithms with exponential speed-up compared to
quantum search that neglects the extra phase shifts.
\end{abstract}



\ifletter\else
\section*{Introduction}
\fi


Quantum algorithms developed by Grover~\cite{grover1997search} and
Shor~\cite{shor1997polynomial} provide the highest motivation for exploring the
computational possibilities offered by quantum mechanics. The crucial
ingredients of the above algorithms, amplitude amplification in the case of
quantum search and phase estimation in the case of quantum factorization, are
important elements of most of the existing quantum
algorithms~\cite{nielsen2010quantum, smith2012algorithms, childs2010quantum}.

In this work we focus on quantum search problems, in particular considering
computation on trees as a problem instance. We aim at developing methods for
harnessing additional knowledge during the search using heuristic methods. In
particular, for broadening the heuristics potential, we try to obtain methods
that allows one to model non-binary evaluation function enabling
multi-threshold approach and ones that do not involve coin operator for
evaluation input.

For the given purpose we choose quantum walk model~\cite{zagury1993walks,
aharonov2001ongraphs, watrous2001undirected} as one of the most promising,
partially due to being universal for quantum
computation~\cite{childs2009universal,PhysRevA.81.042330} and relatively
intuitive. Quantum search with quantum walk is a well studied
field~\cite{portugal2013quantum}. Apart from the basic case a number of
structures has been considered for spatial
search~\cite{childs2004spatialsearch,shenvi2003walksearch,marquezino2013hanoi,konno2013spidernets,PhysRevA.91.062323}.
 Additionally, techniques that allows to perform fixed point 
search~\cite{yoder2014fixed}, with almost arbitrary 
operations~\cite{tulsi2015faster,PhysRevA.91.052307}, fault 
ignorant~\cite{1367-2630-16-7-073033} or a search without an 
oracle~\cite{hillery2012quantum} exists. Quantum search is also considered in 
context of quantum foundations~~\cite{bao2015grover,Tarrataca2012}.

In this work we start with consideration of quantum search on a
tree~\cite{dimcovic2011framework, PhysRevA.77.052104}, which is motivated by
the fact that any connected network can be considered using spanning tree and
tree graphs occur naturally in modelling decision problems.

The model we consider greatly depends on possibility to enable and disable
connections due to its evaluation. Such mechanism is used in many contexts. One
seeks a method to reduce the set of states \eg~by selecting a subspace,
harnessing symmetries, and in consequence obtaining some kind of regularity in
the structure~\cite{novo2014systematic}. Also, studying open systems is often
done by simulating percolation, methods for dynamical control of underlying
graph structure are of interest~\cite{elster2015quantum}. The methods presented
in our work allow one to arbitrarily change connections by changing phase
shifts in the network or its interpretation. We consider disabling one edge in
a pair as restriction of direction according to one of the dimensions. One of
the models where such restriction is present is so called self-avoiding walk
which is an active field of research. In particular tunable self-avoidance is
recently analysed~\cite{camilleri2014quantum}, which may be obtained for a walk
with many different phase shifts with dynamically changing interpretation.

As brute force search complexity is bounded from below, the considered approach
may work only when additional knowledge is accessible. In general it would be
advantageous if quantum heuristics could be applicable for any NP hard
problem~\cite{el2009study}. Harnessing additional knowledge during the walk has
been studied in context of discrete cellular automata, where \eg~feed-forward
coin dependence is modelled~\cite{shikano2014discrete}. Using some concepts of
binary heuristics with tunable threshold in quantum tree search has been shown
to be worth analysing in context of production systems and decision
problems~\cite{Tarrataca2011}. The possibilities of harnessing known heuristics
methods for some class of classical problems, that provide better than
quadratical speed-up and make quantum brute force methods useless in most of
the problem instances has been studied for backtracking
methods~\cite{montanaro2015quantum} as well as a search with
advice~\cite{montanaro2011quantum}. We highlight that the decisions made on the
node during the walk are similar to the ones considered for mobile agents. The
possibility to use agents affecting the measurement settings to improve quantum
network exploration~\cite{miszczak2014magnus} or
search~\cite{tiersch2014adaptive} has been analysed and the authors lay out a
path for adaptive controllers based on intelligent agents for quantum
information tasks.

In this work we highlight the possibility to obtain non-binary evaluation
function. Harnessing many different phase shifts allow many thresholds.
Analysis of quantum inspired ways of multi-thresholding shows that it is
beneficial for image processing~\cite{dey2014multi}.

It is also worth noting that also adiabatic quantum computation model has been
modified in order to harness heuristic guesses~\cite{perdomo2011study}. Joining
quantum and classical techniques into one procedure can also be beneficial even
when the quadratical speed-up is lost by avoiding non-perfect
randomness~\cite{maucher2011search}. The idea can be applied also for purely
quantum decision problems \eg~for oblivious set-member decision
problem~\cite{shi2015two}. More advanced tasks one obtains a decision tree
often may be characterized with use of a decision tree, which would fit into
presented approach.

We combine amplitude amplification with phase estimation to develop a framework
for exploring quantum networks. We introduce a model of quantum network which
allows us to utilize information about the connections. We apply the introduced
framework to develop a quantum walk search procedure exploiting the structural
information in the network. We show that this search procedure allows one to
obtain exponential speed-up over the quantum search procedure which does not
utilize the information available in the network.

\ifletter
\else
This work is organized as follows.
In section \ref{sec:network} the network and the walk model is introduced.
In section \ref{sec:algorithm} we provide the algorithm and discuss the 
complexity.
Section \ref{sec:conclusion} provides concluding remarks.
\fi
\section*{Network model}\label{sec:network}
The results presented in this paper are based on the quantum walk model on a 
network that consists of various types of edges. We aim at encoding information 
about network structure into these types and extracting this knowledge during 
the computation.

\subsection*{Phase altering network}\label{sec:network:model}

Let us consider a network of $n$ nodes with corresponding label set $V$ and
edge set $E$. We assume that the degree of each node is equal to $d$ and
describe the edge labelling by a function $D(x,y)=c\in\{1,\ldots,d\}$ for
$(x,y)\in E$ such that $D(x,y)\ne D(x,y')$. For describing the dynamics of the
system we use a quantum walk
model~\cite{Reitzner,portugal2013quantum,kempe2003overview}. The model
described is consistent with the most general form used for implementing walks
on various structures~\cite{shenvi2003walksearch, douglas2013complexity,
marquezino2013hanoi, sadowski2014spatial} The corresponding quantum system is a
Hilbert space $\XX=\XX_P\otimes\XX_C, \XX_P=\Complex^n, \XX_C=\Complex^d$. Each
basis state $\ket{p,c}=\ket{p}\otimes\ket{c}$ encodes position $p$ and
direction $c$. The most commonly used shift operator takes the form
\begin{equation}
 \sum_{(x,y)\in E} \ket{y, D(y,x)}\bra{x, D(x,y)}.
\end{equation}
However, such model does not differentiate the edges. To achieve this effect we 
introduce a model in which the shift operator includes an edge-dependent phase 
shift
\begin{equation}
S = \sum_{(x,y)\in E} e^{i\varphi(x,y)}\ket{y, D(y,x)}\bra{x, 
D(x,y)}.\label{eq:operator_S}
\end{equation}
In this paper we assume that $\varphi(x,y)=\varphi(y,x)$ so that the phase
marks the edge, not the direction. Additionally, while in general the phases
may be chosen from $[0,2\pi]$, we assume that it is from a narrowed interval,
$\varphi(x,y)<\pi$, so that the two-shift operator $S^2$ exhibits unique phases
$\varphi(x,y)+\varphi(y,x)<2\pi$.

A network defined in such a way enables us to introduce a coin operator that
utilizes the internal structure without using labels assigned to directions.

\subsection*{Phase estimation within quantum walk}\label{sec:network:pahseestim}
We use quantum phase estimation for computing phases resulting from shift
operator application. In order to perform the phase estimation we include an
additional memory register. We extend the Hilbert space with a qudit space
$\XX_M=\Complex^p$ so that the basis states are of the form
$\ket{x,c,m}\in\Complex^n\otimes\Complex^{d}\otimes\Complex^p$. This enables us
to include the auxiliary register for phase estimation and compute the phase
label utilizing the Fourier basis on the memory register with $\mathcal{F}\in
L(\Complex^p)$:
\begin{equation}
\mathcal{F}\ket{k}=\frac{1}{\sqrt{p}}\sum_{j=0}^{p-1} e^{i\varphi_k j} 
\ket{j},
\label{eq:fourier_operator}
\end{equation}
where $\varphi_k=\frac{2\pi k}{p}$ and $k=0,\ldots, p-1$.
We construct the phase estimation operator $E\in 
L(\XX_P\otimes\XX_C\otimes\XX_M)$ using the controlled 
two shifts $S^2$ operator
\begin{equation}
E = (\1_P\otimes\1_C\otimes\mathcal{F}^\dagger)(\sum_{j=0}^{p-1} 
(S^{2})^j\otimes\ketbra{j}{j}).
\label{eq:operator_E}
\end{equation}
In order to implement operator $E$ in a quantum walk scheme without altering 
the shift operator one can introduce auxiliary loops that mimic the desired 
behaviour.

auxiliary direction $c'$ for each direction $c$ such that $c'$ denotes an
inactive counterpart of the $c$ direction ($S\ket{x, c'} =\ket{x, c'}$). We
control number of applications of the shift operator by applying additional
coin operator that at step $l$ deactivates edges with state $\ket{l}$ at the
memory register:
\begin{equation}
\sum_{j=0}^{p-1} (S^{2})^j\otimes\ketbra{j}{j}
=(\1_P\otimes\1_{c\leftrightarrow c'}\otimes\1_M)
\prod_{l=0}^{p-1}\left(
  (S^2\otimes\1_M)D_l\right),
\end{equation}
where we introduce active-inactive switch $\1_{c\leftrightarrow c'} = \1_C - 
\ketbra{c}{c}- \ketbra{c'}{c'}+ 
\ketbra{c'}{c}+ \ketbra{c}{c'}$ and 
controlled deactivation operator
$D_l=\sum_{c=1}^{p-1}\1_P\otimes(
    \1_{c\leftrightarrow c'}\otimes\ketbra{l}{l}+
    \1_C\otimes\sum_{j\ne l}\ketbra{j}{j}$.

The two-shift operator $S^2$ has eigenvectors of the form $\ket{x,y}$ with
eigenvalues $e^{2i\varphi(x,y)}$. Thus the phase estimation procedure $E$
results in
\begin{equation}
E:
\frac{1}{\sqrt{p}}\sum_j\ket{x,y,j}\longmapsto
\ket{x,y}\otimes\ket{\mathrm{ph}(x,y)},
\end{equation}
where $\varphi(x,y)= \frac{2\pi\mathrm{ph}(x,y)}{p}$.
As the result the label ph(x,y) of the phase $\varphi(x,y)$ is encoded into the 
memory register.

\section*{Proof of concept search algorithm}\label{sec:algorithm}
The framework  introduced above can be used to accelerate the execution of
quantum algorithms in networks by utilizing the information about the 
connections. To demonstrate this effect we define a restricted quantum
search problem and provide its solution based on the phase estimation algorithm
performed locally on the nodes. In this case we obtain quadratic speed-up
compared to the quantum search procedure ignoring the phase shifts.

Subsequently we argue that the similar effect can be obtained in more 
general scenarios, in particular in the case of search on a multilevel network, 
where the speed-up rate may be exponential.

\subsection*{Quantum search with prior knowledge}
In this section we consider a search problem where marked node $m$ is to be 
found with use of the oracle operator that flips the phase of the 
corresponding state $\ket{m}$ of the position register: $O_m\ket{m}=-\ket{m}$.
We assume that it is possible to obtain additional information about the 
structure during the walk. In order to maintain quadratical speed-up over 
reduced search space we aim at performing restricted walk harnessing the 
information encoded in the phase shifts and find the marked node using optimal 
number of oracle queries.

\subsection*{Perfect tree case}
A perfect $k$-ary tree is a tree where within each level every node has either 
0 or $k$ children and in which all leaf nodes are at the same depth \ie~a tree 
with constant branching factor.

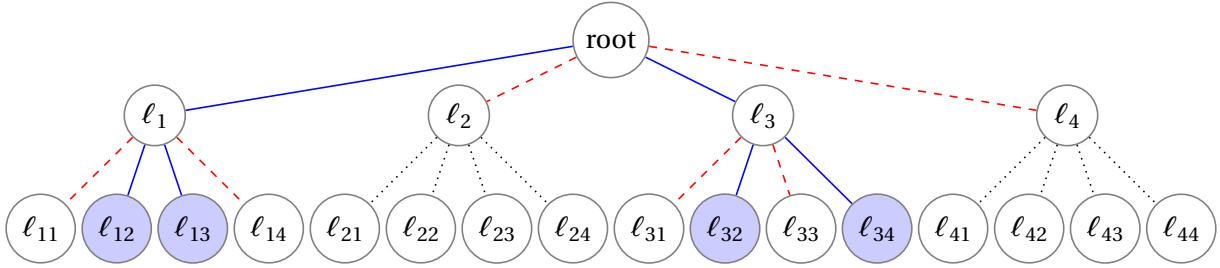
\begin{figure*}[h]
\newlength{\offsetpage}
\setlength{\offsetpage}{2.0cm}
\begin{adjustwidth}{-\offsetpage}{-\offsetpage}%
    \addtolength{\textwidth}{2\offsetpage}
\centering

\begin{tikzpicture}[
auto,
node distance=2cm,
semithick,
bend angle=10,
graybox/.style = {draw=gray!20, fill=gray!20, rounded corners},
line/.style = { draw=black, thick},
box/.style = {circle, draw=black!50,  minimum size=4mm}
]

\coordinate (S) at (0cm, 0cm);

\coordinate (S1) at (-6cm, -1cm);
\coordinate (S2) at (-2cm, -1cm);
\coordinate (S3) at ( 2cm, -1cm);
\coordinate (S4) at ( 6cm, -1cm);

\coordinate (M00) at (-7.5cm, -2.5cm);
\coordinate (M01) at (-6.5cm, -2.5cm);
\coordinate (M02) at (-5.5cm, -2.5cm);
\coordinate (M03) at (-4.5cm, -2.5cm);

\coordinate (M10) at (-3.5cm, -2.5cm);
\coordinate (M11) at (-2.5cm, -2.5cm);
\coordinate (M12) at (-1.5cm, -2.5cm);
\coordinate (M13) at ( -.5cm, -2.5cm);

\coordinate (M20) at ( .5cm, -2.5cm);
\coordinate (M21) at (1.5cm, -2.5cm);
\coordinate (M22) at (2.5cm, -2.5cm);
\coordinate (M23) at (3.5cm, -2.5cm);

\coordinate (M30) at (4.5cm, -2.5cm);
\coordinate (M31) at (5.5cm, -2.5cm);
\coordinate (M32) at (6.5cm, -2.5cm);
\coordinate (M33) at (7.5cm, -2.5cm);

\coordinate (T)  at (4.5cm, -3cm);
\coordinate (T2) at (5.5cm, -3cm);
\coordinate (T3) at (6.5cm, -3cm);

\node (Sbox)  [box] at (S)  {root};

\node (Sbox1) [box] at (S1) {$\ell_{1}$};
\node (Sbox2) [box] at (S2) {$\ell_{2}$};
\node (Sbox3) [box] at (S3) {$\ell_{3}$};
\node (Sbox4) [box] at (S4) {$\ell_{4}$};

\node (Mbox00) [box] at (M00) 				{$\ell_{11}$};
\node[fill=blue!20] (Mbox01) [box] at (M01) {$\ell_{12}$};
\node[fill=blue!20] (Mbox02) [box] at (M02) {$\ell_{13}$};
\node (Mbox03) [box] at (M03) 				{$\ell_{14}$};

\node (Mbox10) [box] at (M10) {$\ell_{21}$};
\node (Mbox11) [box] at (M11) {$\ell_{22}$};
\node (Mbox12) [box] at (M12) {$\ell_{23}$};
\node (Mbox13) [box] at (M13) {$\ell_{24}$};

\node 				(Mbox20) [box] at (M20) {$\ell_{31}$};
\node[fill=blue!20] (Mbox21) [box] at (M21) {$\ell_{32}$};
\node 				(Mbox22) [box] at (M22) {$\ell_{33}$};
\node[fill=blue!20] (Mbox23) [box] at (M23) {$\ell_{34}$};

\node (Mbox30) [box] at (M30) {$\ell_{41}$};
\node (Mbox31) [box] at (M31) {$\ell_{42}$};
\node (Mbox32) [box] at (M32) {$\ell_{43}$};
\node (Mbox33) [box] at (M33) {$\ell_{44}$};

\draw[blue] (Sbox) -- (Sbox1);
\draw[red, dashed] (Sbox) --  (Sbox2);
\draw[blue] (Sbox) --  (Sbox3);
\draw[red, dashed] (Sbox) --  (Sbox4);

\draw[red, dashed] (Sbox1) --  (Mbox00);
\draw[blue] (Sbox1) --  (Mbox01);
\draw[blue] (Sbox1) --  (Mbox02);
\draw[red, dashed] (Sbox1) --  (Mbox03);
\draw[dotted] (Sbox2) --  (Mbox10);
\draw[dotted] (Sbox2) --  (Mbox11);
\draw[dotted] (Sbox2) --  (Mbox12);
\draw[dotted] (Sbox2) --  (Mbox13);

\draw[dotted] (Sbox4) --  (Mbox30);
\draw[dotted] (Sbox4) --  (Mbox31);
\draw[dotted] (Sbox4) --  (Mbox32);
\draw[dotted] (Sbox4) --  (Mbox33);

\draw[red, dashed] 	(Sbox3) --  (Mbox20);
\draw[blue] 		(Sbox3) --  (Mbox21);
\draw[red, dashed] 	(Sbox3) --  (Mbox22);
\draw[blue] 		(Sbox3) --  (Mbox23);

\end{tikzpicture}
\end{adjustwidth}

\caption{Network based on a perfect k-ary tree of degree $k=4$. Blue solid, red 
dashed and black dotted lines represent valid, invalid and unassigned edges  
respectively. The number of valid nodes (filled blue) is reduced 
quadratically.}\label{fig:network}
\end{figure*}

Let us consider a perfect $k$-ary tree network with $k=2d$ and depth $L$ with
$N=(2d)^L$ leaf nodes. We assume that there are two types of edges: valid and
invalid. At each node the possible directions are grouped in $d$ pairs so that
within each pair one direction is marked as valid and one as invalid. Thus
there are $n=d^L=N^{\log d/\log 2d}$ valid nodes \ie~leaf nodes accessible from
the root with valid edges. The example of such setup with $2d=4$ and
$n=\sqrt{N}$ is presented in Figure \ref{fig:network}. We assume that one of
the valid nodes is marked by the oracle operator. The goal is to perform a walk
only on the valid edges and, in result, to reduce the required number of steps.

\begin{algorithm}
\caption{Amplitude amplification with the use of a spreading operator 
$U$ and the number of valid nodes equal to $n$
\label{alg:ampliampli}}
\begin{enumerate}
\item Prepare the initial state $\ket{0}$ at an arbitrary position.
\item Apply operator $U$  to spread on the valid edges.
\item Perform $\ceil{\frac{\pi}{4}\sqrt{n}}$ steps:
\begin{enumerate}
\item Apply the oracle operator.
\item Reverse the spreading with the inverse operator $U^\dagger$.
\item Apply operator $\1-2\ketbra{0}{0}$.
\item Perform the spreading with operator $U$.
\end{enumerate}
\item Measure the position register.
\end{enumerate}
\end{algorithm}

The proposed search algorithm is based on amplitude
amplification~\cite{amplitudeamplification} as described in Algorithm
\ref{alg:ampliampli}. The crucial part of Algorithm \ref{alg:ampliampli} is to
perform the spreading part in such a way that one prepares equal superposition
of the valid leaves only. The method is presented as Algorithm
\ref{alg:spreading}. By combining these procedures the probability of measuring
the marked node is $1/\sqrt{N}$ instead of $1/N$ and the resulting search
complexity is $O(\sqrt[4]{N})$.

\begin{corollary}
For a search problem on a tree network of degree 4 with phase changing edges 
$\varphi\in\{0, \pi/2\}$ grouped in pairs where half of the edges are valid, 
Algorithm \ref{alg:ampliampli} allows finding the marked vertex in 
$O(\sqrt[4]{N})$ steps 
with probability $O(1)$.
\end{corollary}
Proof of the corollary is a consequence of Amplitude amplification properties 
and 
Fact~\ref{fact:spreading}.
\begin{theorem}[Amplitude amplification~\cite{amplitudeamplification}]
Lets define a reflection operator $R_0$ that flips the phase of the initial 
state $\ket{0}$ and does not affect any orthogonal state, an oracle operator 
$O_m$ that flips the phase of the marked state $\ket{m}$ and does not affect 
any orthogonal state and an spreading operator $A$ such that 
$|\bra{m}A\ket{0}|^2=a$.
Suppose $a > 0$, and set $t =\lfloor\pi/4 \theta_a⌋\rfloor$, where $\theta_a$ 
is defined so that $\sin^2(\theta_a) = a$ and $0 < \theta_a \le \pi/2$. Then, 
if 
we compute 
$(AR_0A^{\dagger}O_m)^tA \ket{0}$ and measure the system, the outcome is good 
with probability at 
least $\max(1 - a, a)$
and there exists a quantum algorithm that finds a good solution with certainty 
using a number of applications of $A$ and $A^{\dagger}$ which is in 
$\Theta(1/\sqrt{a})$ in the worst case.
\end{theorem}
Given that the spreading operator $U$ allows one to measure the marked node 
with probability $1/\sqrt{N}$ the overall computational complexity is 
$O(\sqrt[4]{N})$. We provide spreading procedure analysis in the next section.

For the purpose of discussing the speed-up, we assume that gaining the
information encoded in the phase is not possible a priori \ie~a procedure 
allowing one to test whether a node is valid is not provided. Thus, 
the alternative to our approach is a search
without the additional knowledge resulting in the search among $N$ labels with
quantum computational complexity equal to $O(\sqrt{N})$. The number of steps
required for the phase estimation results in a constant factor change in the
complexity. As such the search with the prior knowledge provides a quadratic
speed-up in the considered case over naive quantum search.

\subsection*{Details of Algorithm \ref{alg:spreading}}
In the following section we consider the spreading operator action.

\begin{fact}\label{fact:spreading}
After applying the spreading operator $U$ to the initial state the probability  
of measuring the marked node is equal to $1/n$.
\end{fact}
We justify the fact by showing that each step of the spreading walk prepares 
equal superposition of the valid nodes, and by induction the resulting state is
equal superposition of the valid leaves with measurement probability of each of 
them equal to $1/n$.

\begin{algorithm}
\caption{Spreading part of Algorithm \ref{alg:ampliampli} -- 
operator $U$ --  restricted to the valid edges
\label{alg:spreading}}
\begin{enumerate}
\item Perform for each of the layers of the network:
\begin{enumerate}
\item Prepare the superposition of directions using the coin operator $P$ given 
in Eq. 
(\ref{eq:operator_p}).
\item Estimate phase using the operator $E$ from Eq. (\ref{eq:operator_E}).
\item Prepare the superposition of valid directions using coin operator~$C$ 
given in Eq. (\ref{eq:operator_C}).
\item Apply the shift operator $S$, given in Eq. \ref{eq:operator_S}, in order 
to transfer onto the lower layer 
with the use of the valid edges.
\item Reset the superposition getting the initial form of the state with 
operator $A$ in Eq. (\ref{eq:operator_A}).
\end{enumerate}
\item Resume the overall algorithm.
\end{enumerate}
\end{algorithm}

We consider a system consisting of three registers: 
position $\XX_P=\Complex^{N(L+1)}$, coin (direction) $\XX_C=\Complex^{d+1}$ and 
memory 
$\XX_M=\Complex^p$. We start in the state

\begin{equation}
\ket{\psi_0} = \ket{0,0,0} \in \XX_P\otimes\XX_C\otimes\XX_M.
\end{equation}
Let us note that in such system each node has the same number of internal 
states corresponding to directions. For the root node and the leaves this does 
not accurately model the tree structure. In order to apply regular graph walk 
model we assign loops to all of the problematic directions.

At each step we begin by preparing the superposition over possible 
directions and initializing the memory in the equal superposition of all 
possible phase label values.
This is implemented by an operator $P\in L(\XX)$ 
that rotates a state in the initial form $\ket{0,0}\in\XX_C\otimes\XX_M$ into 
$\ket{s}=\sum_{c}\sum_{j}\ket{c,j}$ at any position $x$:
\begin{equation}
P = \sum_x \ketbra{x}{x}\otimes\1_{\ket{0,0}\rightarrow \ket{s}}.
\label{eq:operator_p}
\end{equation}
Operator $U_{\ket{a}\rightarrow \ket{b}}$ alters only states from
$\mathrm{span}(\ket{a}, \ket{b})$
with operator $\ketbra{b}{a} + \ketbra{b'}{a'}$.
In particular
\begin{equation}
U_{\ket{a}\rightarrow \ket{b}}\ket{x}=\left\{\begin{matrix}
\ket{x},& \ket{x}\bot\mathrm{span}(\ket{a}, \ket{b}),\\
\ket{b},& \ket{x}=\ket{a},\\
\ket{b'},& \ket{x}=\ket{a'},
\end{matrix}\right.
\end{equation}
where $\ket{a'}$ and $\ket{b'}$ 
denote normalized states $\ket{b}-\ket{a}\braket{a}{b}$ and 
$\ket{a}-\ket{b}\braket{b}{a}$ respectively.

Then we perform the phase estimation procedure given in Eq.
(\ref{eq:operator_E}). The algorithm uses the memory register to control the
overall phase. After performing the Fourier transform we obtain the value of
the phase at the memory register.

Having determined which phase is assigned to which direction we transform equal
superposition of all of the directions into a superposition of the valid
directions only. Due to the grouping of the directions and phases we can
perform this transition with the use of a unitary operator $C\in
L(\XX_C\otimes\XX_M)$ acting on joint coin and memory registers:

\begin{equation}
\begin{split}
C=\frac{1}{\sqrt{2}}\sum_{l=0}^{d-1}&
\ket{2l,g}\bra{g_{2l}^+}+\ket{2l+1,g}\bra{g_{2l+1}^+}+\\&
\ket{2l+1,b}\bra{g_{2l+1}^-}+\ket{2d,b}\bra{g_{2l}^-},
\end{split}\label{eq:operator_C}
\end{equation}
where the $g,b$ stand for valid and invalid phase labels. The introduced
operator is designed to properly transform a state of a pair $l$ of directions
$2l, 2l+1$ depending on which direction is valid. If direction $2l$  (or
$2l+1$) is valid the state of the pair is
$\ket{g_{2l}^+}=\ket{2l,g}+\ket{2l+1,b}$ (or
$\ket{g_{2l+1}^+}=\ket{2l,b}+\ket{2l+1,g}$ correspondingly) and is transformed
into a valid direction $\ket{2l, g}$ (or $\ket{2l+1,g}$). One should note that
the restriction (directions grouped in pairs) is introduced in order to provide
a simple unitary example useful for a search task. For other tasks, such as
routing, where the spreading does not have to be reversible, it is possible to
implement the direction selection, with more general quantum channel using open 
walk model, and avoid such limitations~\cite{pawela2014generalized}.

When valid directions are encoded in the coin register we perform the shift 
operator and then transform resulting states into the initial form with  
operator

\begin{equation}
A = \sum_x \ketbra{x}{x}\otimes
\1_{\ket{\uparrow,g}\rightarrow\ket{0,0}},
\label{eq:operator_A}
\end{equation}
where $\uparrow$ denotes the direction leading to the parent node and $g$ the 
valid phase label.

As the final result we obtain equal superposition over all leaves that are
reachable using only valid edges. This enables us to use procedure described in
Algorithm \ref{alg:ampliampli} to execute search on valid nodes only.

\subsection*{Generalized tree case}\label{sec:algorithm:general}
In the example presented above we have restricted ourselves to the situation
where only two types of edges are distributed equally in the network layers. 
However, the presented scheme can be applied to more complicated phase
configurations and validation methods.
It is important to note that the setting is straightforward to generalize in
order to increase the number of edges in the group and in result decrease the
factor of valid nodes.

In general if only 2 out of $2^R$
(in general $k$ out of $k^R$) of the edges are valid at each node the number of
valid nodes is reduced accordingly: $n=\sqrt[R]{N}$. We obtain search
complexity of order $O(\sqrt{n})$, compared to $O((\sqrt{n})^R)$,  This class
of problems generates arbitrarily large polynomial speed-up. The polynomial
degree increases linearly with with exponential growth of the degree of the
tree and corresponding valid edges ratio.

The grouping of the edges assumed in the model of unitary selection of the
valid edges can be seen as clustering separate dimensions.
In such scenario we restrict the walker to move only in one direction within
the dimension. Such behaviour commonly occurs in quantum walks. When
considering networks with simple representation in the Fourier basis the edges
are naturally grouped into classes~\cite{ambainis2005coins}. A similar approach
also frequently appears in scale free networks and other regular graph-based
models~\cite{marquezino2013hanoi}.

\subsection*{Multilevel network case}
\def \Vbottomlevel {-3cm}
\def \bottomlevel {-5cm}
\def \Pbottomlevel {-7cm}

The tree network model discussed in the previous section is suitable to show
the basic example in details, but it can only exhibit polynomial 
complexity speed-up rate. In more general cases additional information can 
cause exponential complexity decay.

The walk model for the multilevel network is analogous to the tree case.
The shift operator act as described in eq. (\ref{eq:operator_S}) taking into 
account modified connections. Operations on the nodes are local and thus almost 
the same. Main difference lies in the fact that instead of having one parent 
node, one need to consider a number of parents and a corresponding state being 
a superposition of proper directions.

\begin{figure*}[t]
\footnotesize

\setlength{\offsetpage}{2.cm}
\begin{adjustwidth}{-\offsetpage}{-\offsetpage}%
    \addtolength{\textwidth}{2\offsetpage}
\centering
\begin{tikzpicture}[
auto,
node distance=2cm,
semithick,
bend angle=10,
graybox/.style = {draw=gray!20, fill=gray!20, rounded corners},
line/.style = { draw=black, thick},
box/.style = {circle, draw=black!50,  minimum size=4mm}
]

\coordinate (S) at (0cm, 0cm);

\coordinate (M00) at (-7.5cm, \bottomlevel);
\coordinate (M01) at (-6.5cm, \bottomlevel);
\coordinate (M02) at (-5.5cm, \bottomlevel);
\coordinate (M03) at (-4.5cm, \bottomlevel);

\coordinate (M10) at (-3.5cm, \bottomlevel);
\coordinate (M11) at (-2.5cm, \bottomlevel);
\coordinate (M12) at (-1.5cm, \bottomlevel);
\coordinate (M13) at ( -.5cm, \bottomlevel);

\coordinate (M20) at ( .5cm, \bottomlevel);
\coordinate (M21) at (1.5cm, \bottomlevel);
\coordinate (M22) at (2.5cm, \bottomlevel);
\coordinate (M23) at (3.5cm, \bottomlevel);

\coordinate (M30) at (4.5cm, \bottomlevel);
\coordinate (M31) at (5.5cm, \bottomlevel);
\coordinate (M32) at (6.5cm, \bottomlevel);
\coordinate (M33) at (7.5cm, \bottomlevel);

\coordinate (V00) at (-7.5cm, \Vbottomlevel);
\coordinate (V01) at (-6.5cm, \Vbottomlevel);
\coordinate (V02) at (-5.5cm, \Vbottomlevel);
\coordinate (V03) at (-4.5cm, \Vbottomlevel);

\coordinate (V10) at (-3.5cm, \Vbottomlevel);
\coordinate (V11) at (-2.5cm, \Vbottomlevel);
\coordinate (V12) at (-1.5cm, \Vbottomlevel);
\coordinate (V13) at ( -.5cm, \Vbottomlevel);

\coordinate (V20) at ( .5cm, \Vbottomlevel);
\coordinate (V21) at (1.5cm, \Vbottomlevel);
\coordinate (V22) at (2.5cm, \Vbottomlevel);
\coordinate (V23) at (3.5cm, \Vbottomlevel);

\coordinate (V30) at (4.5cm, \Vbottomlevel);
\coordinate (V31) at (5.5cm, \Vbottomlevel);
\coordinate (V32) at (6.5cm, \Vbottomlevel);
\coordinate (V33) at (7.5cm, \Vbottomlevel);

\coordinate (P00) at (-7.5cm, \Pbottomlevel);
\coordinate (P01) at (-6.5cm, \Pbottomlevel);
\coordinate (P02) at (-5.5cm, \Pbottomlevel);
\coordinate (P03) at (-4.5cm, \Pbottomlevel);

\coordinate (P10) at (-3.5cm, \Pbottomlevel);
\coordinate (P11) at (-2.5cm, \Pbottomlevel);
\coordinate (P12) at (-1.5cm, \Pbottomlevel);
\coordinate (P13) at ( -.5cm, \Pbottomlevel);

\coordinate (P20) at ( .5cm, \Pbottomlevel);
\coordinate (P21) at (1.5cm, \Pbottomlevel);
\coordinate (P22) at (2.5cm, \Pbottomlevel);
\coordinate (P23) at (3.5cm, \Pbottomlevel);

\coordinate (P30) at (4.5cm, \Pbottomlevel);
\coordinate (P31) at (5.5cm, \Pbottomlevel);
\coordinate (P32) at (6.5cm, \Pbottomlevel);
\coordinate (P33) at (7.5cm, \Pbottomlevel);

\node[fill=blue!20] (Vbox00) [box] at (V00) {};
\node[fill=blue!20] (Vbox01) [box] at (V01) {};
\node[fill=blue!20] (Vbox02) [box] at (V02) {};
\node[fill=blue!20] (Vbox03) [box] at (V03) {};
\node[fill=blue!20] (Vbox10) [box] at (V10) {};
\node[fill=blue!20] (Vbox11) [box] at (V11) {};
\node[fill=blue!20] (Vbox12) [box] at (V12) {};
\node[fill=blue!20] (Vbox13) [box] at (V13) {};
\node[fill=blue!20] (Vbox20) [box] at (V20) {};
\node[fill=blue!20] (Vbox21) [box] at (V21) {};
\node[fill=blue!20] (Vbox22) [box] at (V22) {};
\node[fill=blue!20] (Vbox23) [box] at (V23) {};
\node[fill=blue!20] (Vbox30) [box] at (V30) {};
\node[fill=blue!20] (Vbox31) [box] at (V31) {};
\node[fill=blue!20] (Vbox32) [box] at (V32) {};
\node[fill=blue!20] (Vbox33) [box] at (V33) {};

\node 				(Mbox00) [box] at (M00) {};
\node[fill=blue!20] (Mbox01) [box] at (M01) {};
\node[fill=blue!20] (Mbox02) [box] at (M02) {};
\node 				(Mbox03) [box] at (M03) {};
\node 				(Mbox10) [box] at (M10) {};
\node[fill=blue!20] (Mbox11) [box] at (M11) {};
\node[fill=blue!20] (Mbox12) [box] at (M12) {};
\node 				(Mbox13) [box] at (M13) {};
\node 				(Mbox20) [box] at (M20) {};
\node[fill=blue!20] (Mbox21) [box] at (M21) {};
\node 				(Mbox22) [box] at (M22) {};
\node[fill=blue!20] (Mbox23) [box] at (M23) {};
\node 				(Mbox30) [box] at (M30) {};
\node[fill=blue!20] (Mbox31) [box] at (M31) {};
\node[fill=blue!20] (Mbox32) [box] at (M32) {};
\node 				(Mbox33) [box] at (M33) {};

\node 				(Pbox00) [box] at (P00) {};
\node 				(Pbox01) [box] at (P01) {};
\node 				(Pbox02) [box] at (P02) {};
\node 				(Pbox03) [box] at (P03) {};
\node[fill=blue!20] (Pbox10) [box] at (P10) {};
\node 				(Pbox11) [box] at (P11) {};
\node[fill=blue!20] (Pbox12) [box] at (P12) {};
\node 				(Pbox13) [box] at (P13) {};
\node 				(Pbox20) [box] at (P20) {};
\node[fill=blue!20] (Pbox21) [box] at (P21) {};
\node 				(Pbox22) [box] at (P22) {};
\node[fill=blue!20] (Pbox23) [box] at (P23) {};
\node 				(Pbox30) [box] at (P30) {};
\node 				(Pbox31) [box] at (P31) {};
\node 				(Pbox32) [box] at (P32) {};
\node 				(Pbox33) [box] at (P33) {};

\draw[blue] 		(Vbox00) --  (Mbox01); 
\draw[blue] 		(Vbox00) --  (Mbox02); 
\draw[blue] 		(Vbox01) --  (Mbox02); 
\draw[blue] 		(Vbox01) --  (Mbox12); 
\draw[blue] 		(Vbox02) --  (Mbox01); 
\draw[blue] 		(Vbox02) --  (Mbox12); 
\draw[blue] 		(Vbox03) --  (Mbox11); 
\draw[blue] 		(Vbox03) --  (Mbox12); 
\draw[blue] 		(Vbox10) --  (Mbox02); 
\draw[blue] 		(Vbox10) --  (Mbox11); 
\draw[blue] 		(Vbox11) --  (Mbox01); 
\draw[blue] 		(Vbox11) --  (Mbox11); 
\draw[blue] 		(Vbox12) --  (Mbox01); 
\draw[blue] 		(Vbox12) --  (Mbox11); 
\draw[blue] 		(Vbox13) --  (Mbox12); 
\draw[blue] 		(Vbox13) --  (Mbox02); 

\draw[gray, dashed] 	(Vbox00) --  (Mbox00); 
\draw[gray, dashed] 	(Vbox00) --  (Mbox03); 
\draw[gray, dashed] 	(Vbox01) --  (Mbox00); 
\draw[gray, dashed] 	(Vbox01) --  (Mbox03); 
\draw[gray, dashed] 	(Vbox02) --  (Mbox10); 
\draw[gray, dashed] 	(Vbox02) --  (Mbox20); 
\draw[gray, dashed] 	(Vbox03) --  (Mbox10); 
\draw[gray, dashed] 	(Vbox03) --  (Mbox20); 
\draw[gray, dashed] 	(Vbox10) --  (Mbox00); 
\draw[gray, dashed] 	(Vbox10) --  (Mbox03); 
\draw[gray, dashed] 	(Vbox11) --  (Mbox00); 
\draw[gray, dashed] 	(Vbox11) --  (Mbox03); 
\draw[gray, dashed] 	(Vbox12) --  (Mbox10); 
\draw[gray, dashed] 	(Vbox12) --  (Mbox20); 
\draw[gray, dashed] 	(Vbox13) --  (Mbox10); 
\draw[gray, dashed] 	(Vbox13) --  (Mbox20); 

\draw[blue] 		(Vbox20) --  (Mbox21); 
\draw[blue] 		(Vbox20) --  (Mbox32); 
\draw[blue] 		(Vbox21) --  (Mbox31); 
\draw[blue] 		(Vbox21) --  (Mbox23); 
\draw[blue] 		(Vbox22) --  (Mbox21); 
\draw[blue] 		(Vbox22) --  (Mbox23); 
\draw[blue] 		(Vbox23) --  (Mbox21); 
\draw[blue] 		(Vbox23) --  (Mbox23); 
\draw[blue] 		(Vbox30) --  (Mbox31); 
\draw[blue] 		(Vbox30) --  (Mbox23); 
\draw[blue] 		(Vbox31) --  (Mbox21); 
\draw[blue] 		(Vbox31) --  (Mbox32); 
\draw[blue] 		(Vbox32) --  (Mbox31); 
\draw[blue] 		(Vbox32) --  (Mbox32); 
\draw[blue] 		(Vbox33) --  (Mbox31); 
\draw[blue] 		(Vbox33) --  (Mbox32); 

\draw[gray, dashed] 	(Vbox20) --  (Mbox13); 
\draw[gray, dashed] 	(Vbox20) --  (Mbox22); 
\draw[gray, dashed] 	(Vbox21) --  (Mbox13); 
\draw[gray, dashed] 	(Vbox21) --  (Mbox22); 
\draw[gray, dashed] 	(Vbox22) --  (Mbox30); 
\draw[gray, dashed] 	(Vbox22) --  (Mbox33); 
\draw[gray, dashed] 	(Vbox23) --  (Mbox30); 
\draw[gray, dashed] 	(Vbox23) --  (Mbox33); 
\draw[gray, dashed] 	(Vbox30) --  (Mbox13); 
\draw[gray, dashed] 	(Vbox30) --  (Mbox22); 
\draw[gray, dashed] 	(Vbox31) --  (Mbox13); 
\draw[gray, dashed] 	(Vbox31) --  (Mbox22); 
\draw[gray, dashed] 	(Vbox32) --  (Mbox30); 
\draw[gray, dashed] 	(Vbox32) --  (Mbox33); 
\draw[gray, dashed] 	(Vbox33) --  (Mbox30); 
\draw[gray, dashed] 	(Vbox33) --  (Mbox33); 

\draw[blue] 		(Mbox01) --  (Pbox10); 
\draw[blue] 		(Mbox01) --  (Pbox12); 
\draw[blue] 		(Mbox02) --  (Pbox10); 
\draw[blue] 		(Mbox02) --  (Pbox12); 
\draw[blue] 		(Mbox11) --  (Pbox21); 
\draw[blue] 		(Mbox11) --  (Pbox10); 
\draw[blue] 		(Mbox12) --  (Pbox12); 
\draw[blue] 		(Mbox12) --  (Pbox10); 
\draw[blue] 		(Mbox21) --  (Pbox21); 
\draw[blue] 		(Mbox21) --  (Pbox23); 
\draw[blue] 		(Mbox23) --  (Pbox21); 
\draw[blue] 		(Mbox23) --  (Pbox23); 
\draw[blue] 		(Mbox31) --  (Pbox23); 
\draw[blue] 		(Mbox31) --  (Pbox12); 
\draw[blue] 		(Mbox32) --  (Pbox23); 
\draw[blue] 		(Mbox32) --  (Pbox21); 

\draw[gray, dashed] 	(Mbox01) --  (Pbox00); 
\draw[gray, dashed] 	(Mbox01) --  (Pbox03); 
\draw[gray, dashed] 	(Mbox02) --  (Pbox11); 
\draw[gray, dashed] 	(Mbox02) --  (Pbox13); 
\draw[gray, dashed] 	(Mbox11) --  (Pbox00); 
\draw[gray, dashed] 	(Mbox11) --  (Pbox02); 
\draw[gray, dashed] 	(Mbox12) --  (Pbox11); 
\draw[gray, dashed] 	(Mbox12) --  (Pbox13); 

\draw[gray, dashed] 	(Mbox21) --  (Pbox20); 
\draw[gray, dashed] 	(Mbox21) --  (Pbox22); 
\draw[gray, dashed] 	(Mbox23) --  (Pbox31); 
\draw[gray, dashed] 	(Mbox23) --  (Pbox33); 
\draw[gray, dashed] 	(Mbox31) --  (Pbox20); 
\draw[gray, dashed] 	(Mbox31) --  (Pbox22); 
\draw[gray, dashed] 	(Mbox32) --  (Pbox30); 
\draw[gray, dashed] 	(Mbox32) --  (Pbox33); 
\end{tikzpicture}
\end{adjustwidth}
\caption{Network based on a multilevel structure. Blue solid and grey dashed 
lines represent valid and invalid edges respectively. Edges outgoing from 
invalid nodes are not shown. The number of valid nodes (filled 
blue) at the bottom level is reduced quadratically.}\label{fig:network2}
\end{figure*}
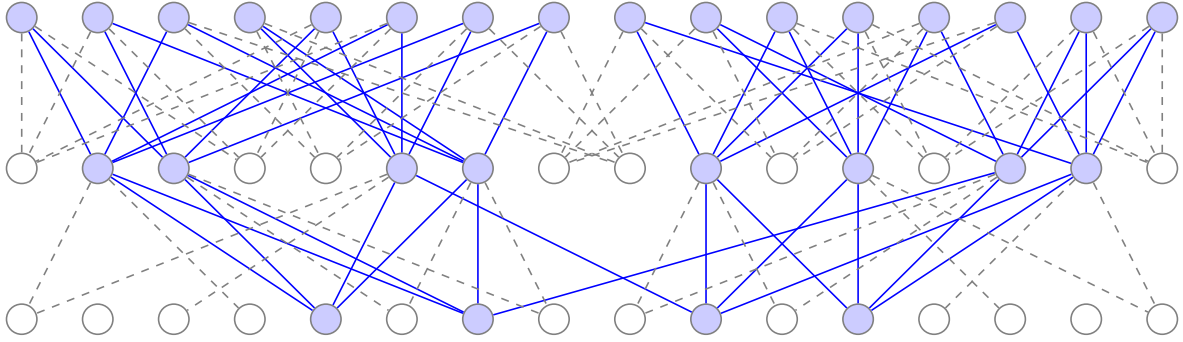

Let us consider a network composed of many layers, such that every layer
consists of the same number of nodes $N$. We assume, that we are able to
prepare a state being an equal superposition of all of the nodes from the top
layer (\eg~with use of a common root-parent node). Again we assume, that there
are 2 pairs of edges outgoing downwards from each of the nodes. Exactly one
edge in every pair is valid. An example of valid edges assignment is shown in
Fig. \ref{fig:network2}.

In this case the measure of the problem size is the number of possible outcomes
$N$, which is equal to the number of nodes in the bottom layer. We consider a
scheme where the number of valid nodes $n$ in the last layer is fixed. Thus,
layer width $N$ grows exponentially as $N=n2^{D-1}$, where $D$ is network depth
(layers number). We are interested in complexity of finding a marked node for
fixed number of valid nodes when the number of layers increases. The quantum
algorithm is very similar to the previous case and the resulting query
complexity is constant for fixed valid nodes set size. The number of required
steps in the corresponding algorithm increases logarithmically.
 
\subsection*{Model potential}
This examples are designed to show the advantages of including the phase shift 
in a walk scheme. In described cases obtained speed-up is always quadratical 
compared to the classical case where all the information encoded into network 
is available, thus the speed-up rate is not better than in the standard 
Grover's search. However, a convenient way to harness classical heuristics in 
quantum search is of great interest, because without the method for 
incorporating the additional knowledge the quantum speed-up gain is lost.
The presented model is especially convenient because of the fact that labelling 
edges in the network with phase shifts does not require any dependence on 
directions labels. Moreover, using greater number of possible phases one could 
describe many different edges classes and change its behaviour dynamically with 
change of the coin operator only. In particular when different edges classes 
correspond to different heuristics evaluation one can control the threshold 
that determine active/inactive edges on the run without altering the network 
structure.

\section*{Concluding remarks}\label{sec:conclusion}

The aim of this work is to study possibilities of developing the quantum walk
model in order to increase its algorithmic potential. In this work we provide
the evidence that the model allows one to perform efficient restricted walk
propagation on tree-alike structures. Moreover, the restrictions can utilize
network structure containing information about the connections without
reference to the directions labelling. Additionally, the information may take a
form of non-binary evaluation function. The further research will focus on the
issue whether the model is suitable for more sophisticated protocols with the
use of similar routing scheme. The presented work gives foundations for
studying computation methods that allows local determination of network
properties and driving the walk behaviour. In particular, the algorithm may be
useful in the context of quantum heuristics and mobile agents methods.

\newif\ifthankncn
\thankncntrue

\section*{Acknowledgements}
The author would like to thank J. A. Miszczak for useful comments and
discussions during the preparation of this paper. This work is supported by the
Polish Ministry of Science and Higher Education within "Diamond Grant"
Programme under the project number 0064/DIA/2013/42 and by the Polish National
Science Centre under the research project 2011/03/D/ST6/00413.


%
%
\end{document}